\documentclass[twocolumn,aps,pre,lengthcheck]{revtex4}
\usepackage[dvips]{graphicx}
\usepackage{amssymb}
\usepackage{amsmath}
\usepackage{latexsym}
\usepackage{epsfig}
\usepackage{bm}
\usepackage{color}
\usepackage{times,psfrag,subfigure}      

\begin{document}
\title{Free energy pathways of a Multistable Liquid Crystal Device}
\author{Halim Kusumaatmaja}
\email{halim.kusumaatmaja@durham.ac.uk}
\affiliation{
Department of Physics, University of Durham, South Road, DH1 3LE, U.K.}
\author{Apala Majumdar}
\affiliation{
Department of Mathematical Sciences, University of Bath, Bath, BA2 7AY, U.K.}
\date{\today}


\begin{abstract}
The planar bistable device [Tsakonas \textit{et al., Appl. Phys. Lett.}, 2007, {\textbf{90}}, 111913] is known to have two distinct classes of stable equilibria: the diagonal and rotated solutions. We model this device within the two-dimensional Landau-de Gennes theory, with a surface potential and without any external fields. We systematically compute a special class of transition pathways, referred to as minimum energy pathways, between the stable equilibria that provide new information about how the equilibria are connected in the Landau-de Gennes free energy landscape. These transition pathways exhibit an intermediate transition state, which is a saddle point of the Landau-de Gennes free energy. We numerically compute the structural details of the transition states, the optimal transition pathways and the free energy barriers between the equilibria, as a function of the surface anchoring strength. For strong anchoring, the transition pathways are mediated by defects whereas we get defect-free transition pathways for moderate and weak anchoring. In the weak anchoring limit, we recover a cusp catastrophe situation for which the rotated state acts as a transition state connecting two different diagonal states.
\end{abstract}
\maketitle


\section{Introduction}
Nematic liquid crystals are complex anisotropic liquids with long-range orientational ordering~\cite{deGennes_book}. Nematics in confinement present a whole host of new theoretical and applications-oriented questions focussed on the complex inter-relationship between material properties, geometry, temperature, boundary effects and external fields in equilibrium and non-equilibrium phenomena. From a modelling point of view, we have new and challenging questions on pattern formation, defects, interfacial phenomena and rheology. From a technological point of view, nematics in micron-scale or nano-scale geometries open new doors for optical and display applications, and more recently novel biological insight~\cite{Lagerwall_2012, Lacaze_2013}.

We re-visit the planar bistable device first reported by Tsakonas and his co-workers~\cite{Tsakonas_2007}. The planar bistable device has been well-studied in the liquid crystal community in recent years \cite{Tsakonas_2007,Majumdar_2012, Kralj_2014, Cleaver_2010, Cleaver_2012}, partly because of its geometrical simplicity and partly because of its rich modelling landscape. It typically comprises a periodic array of shallow square or rectangular wells filled with nematic liquid crystalline material. The well surfaces are treated to induce tangent or planar boundary conditions so that the nematic molecules, in contact with the well surfaces, are constrained to be tangent to the well surfaces. Tsakonas et al.~\cite{Tsakonas_2007} adopt a two-dimensional Landau-de Gennes modelling approach and focus on planar nematic equilibria on the bottom square or rectangular well cross-section, on the grounds that non-planar three-dimensional equilibria are energetically expensive for shallow three-dimensional wells and hence, not physically relevant. They compute local minimizers of a two-dimensional Landau-de Gennes energy on a square or a rectangle with tangent boundary conditions and obtain two distinct classes of optically contrasting nematic equilibria: the diagonal solutions and the rotated solutions and their modelling results are supported by parallel experimental work. The
tangent boundary conditions naturally create a mismatch in molecular alignments or defects at the square vertices. As the name suggests, the nematic molecules roughly align along the square diagonal for the diagonal solution. The rotated solutions have a more distorted profile as simulations and optical data suggest that the molecules rotate by about 180 degrees between a pair of parallel square edges. There are two diagonal solutions, one for each square diagonal, and four rotated solutions, related to each other by a 90 degree rotation. Luo {\it{et al.}}~\cite{Majumdar_2012} build on the work in \cite{Tsakonas_2007} and carefully study the dependence of the diagonal and rotated solutions on the surface anchoring strength, denoted by a surface anchoring coefficient $W$. In particular, they find that the rotated solutions only exist above a certain critical anchoring strength and the critical anchoring strongly depends on the material parameters and temperature.

Here we take the modelling work further by a systematic numerical investigation of the free energy pathways of the planar bistable device within the framework of the Landau-deGennes theory. This is a significant forward step since we not only recover the free energy minimizing states but also compute structural and energetic information about the transient states connecting pairs of distinct free energy minima. As we vary the surface anchoring strength, we find three distinct regimes: the strong anchoring, moderate anchoring and weak anchoring regimes. In the strong anchoring limit, we recover the familiar diagonal and rotated solutions as local free energy minimizers. As noted in \cite{Tsakonas_2007,Majumdar_2012}, there are two different diagonal solutions and four distinct rotated solutions. We compute the transition pathways between the different stable solutions. In particular, we focus on the so-called minimum energy pathways, where every point on such a pathway is an energy minimum in all but a single distinguished direction in the phase space. There are a number of interesting findings to highlight here. Firstly, each minimum energy pathway is featured by a transition state, which is a saddle-point, an unstable critical point of the Landau-de Gennes energy. We identify the transition states as being local free energy maxima along a minimum energy pathway connecting free energy minima. In the strong anchoring limit, the transition states exhibit $\pm 1/2$ defects along the square edges. There may be multiple minimum energy pathways between a pair of distinct minima and these transition pathways can contain different number of defects. We interpret the optimal transition pathway as being the minimum energy pathway with the smallest free energy barrier and the least energetically expensive transition state. It is noteworthy that we do not find a direct minimum energy pathway between pairs of diagonal solutions or pairs of rotated solutions and all minimum energy pathways connect diagonal and rotated solutions. In the moderate anchoring regime, the transition pathways are not mediated by defects but are rather mediated by localized anchoring breaking along the edges, which induces a global transition between diagonal and rotated solutions. In the weak anchoring regime, the rotated solutions cease to be locally stable (consistent with the numerical findings in \cite{Majumdar_2012}) but do exist as transition states connecting the diagonal solutions. In particular, the numerical methods in \cite{Tsakonas_2007,Majumdar_2012} cannot capture the persistence of rotated solutions, as non energy-minimizing solutions, in the weak anchoring regime.

We make some remarks on the novelty and importance of our approach. The transition states are not stable but they may be experimentally observable. Indeed the transition states in the strong anchoring limit are reminiscent of recent experimental observations of metastable states with point defects in shallow nematic chambers \cite{Aarts_2014}. Further, our numerical results demonstrate a subtle dependence of the optimal transition pathways on the anchoring strength. In the strong anchoring limit, the optimal transition pathway is almost independent of the anchoring strength whereas the free energy barrier decreases monotonically with anchoring strength in the moderate and weak anchoring regimes. Switching mechanisms rely on a complex interplay between anchoring strength, material properties and external fields and one could use our numerical methods to compute and analyze optimal switching pathways 
in the presence of external fields. 

The paper is organized as follows. In Section~\ref{sec:2}, we review the Landau-de Gennes theory for nematic liquid crystals. In Section~\ref{sec:3}, we compute the Landau-de Gennes free energy minimizers and in Section~\ref{sec:4}, we compute the transition states and the transition pathways between free energy minima, focussing on three different anchoring regimes. We conclude in Section~\ref{sec:5} with future perspectives. The computational methodologies used in this paper are described in the appendix.

\section{Computational Model}
\label{sec:2}

Following the paradigm in \cite{Tsakonas_2007, Majumdar_2012}, we model the planar bistable device within the two-dimensional Landau-de Gennes theory. In Landau-de Gennes theory, all the microscopic details about molecular shape and interactions are averaged out, and the nematic state is described by an macroscopic order parameter, ${\bf{Q}}$. We take the computational domain, $\Omega$, to be a square in the $(x,y)$-plane, which defines the bottom surface of the well,
\begin{equation}
\label{eq:1}
\Omega =  \left\{(x,y) \in {\it{\bf{R}}}^2: 0\leq x,y \leq L \right\},
\end{equation} where $L$ denotes the system size. As in \cite{Majumdar_2012}, for two dimensions, the ${\bf{Q}}$ tensor can be represented by a symmetric traceless $2 \times 2$ matrix,
\begin{equation}
{\bf{Q}} = \begin{bmatrix}
       Q_{11} & Q_{12}           \\
       Q_{12} & -Q_{11}
     \end{bmatrix}
= s(2\bf{n}\otimes\bf{n}-I), \label{Qtensor}
\end{equation}
where $\bf{n}$ is the director or the distinguished direction of molecular alignment and $s$ is a scalar order parameter that measures the degree of orientational ordering about $\bf{n}$.

We work with a simple form of the Landau-de Gennes energy as given below, with no external fields.
\begin{eqnarray}
\Psi &=& \int_\Omega \left[-\alpha\mathrm{Tr}{\bf{Q}}^2-\frac{B}{3}\mathrm{Tr}{\bf{Q}}^3+\frac{C}{4}(\mathrm{Tr}{\bf{Q}}^2)^2\right] dA  \label{feLC} \\
&+&\int_\Omega \frac{\kappa_{el}}{2} |\nabla {\bf{Q}}|^2 dA + \int_{\partial A} W |(Q_{11},Q_{12})-{\bf{q}}|^2 da. \nonumber
\end{eqnarray}
The first integral is the bulk free energy that drives the nematic-isotropic transition as a function of the temperature \cite{deGennes_book,Virga_book}. The coefficient $\alpha$ is the re-scaled temperature; we work with temperatures below the critical nematic supercooling temperature
and hence, $\alpha>0$; the coefficients $B, C$ are positive material-dependent constants. For a two-dimensional $\mathbf{Q}$ as in (\ref{Qtensor}), $\textrm{tr}\mathbf{Q}^3 = 0$, and the bulk free energy simplifies to the familiar Ginzburg-Landau potential \cite{Majumdar_2012}. The second term is the one-constant elastic energy and $\kappa_{el} > 0$ is an elastic
constant; there are more general quadratic elastic energy densities but we believe that the one-constant energy density suffices for qualitative purposes. The third term is a Durand-Nobili surface anchoring energy~\cite{Nobili_1992} that enforces the preferred tangential anchoring on the square edges, with anchoring strength $W$, reference configuration $\mathbf{q}$ and boundary element $da$. There are multiple choices for the surface energy (see \cite{Majumdar_2012} for comparisons between three potential candidates for the tangent surface energy) but the Durand-Nobili energy has the desired numerical stability properties for all relevant ranges
of $W$.

To reduce the number of input parameters in the model, we now rewrite the free energy functional in its dimensionless form. Substituting Eq. \eqref{Qtensor} into \eqref{feLC} and defining $\tilde{x}=x/L$, $\tilde{\Psi} =C\Psi/\alpha^2L^2$, $\tilde{\kappa}_{el} =
\kappa_{el}/\alpha L^2$, $\tilde{\mathbf{Q}}^2 = C \mathbf{Q}^2/\alpha$, $\tilde{W} = W /\alpha L$, we obtain

\begin{eqnarray}
\tilde{\Psi} &=& \int_{\tilde{A}} (\tilde{Q}_{11}^2+\tilde{Q}_{12}^2-1)^2  d\tilde{A}
\nonumber \\
&+& \int_{\tilde{A}} \tilde{\kappa}_{el} \left[|\tilde{\nabla} \tilde{Q}_{11}|^2 +  |\tilde{\nabla} \tilde{Q}_{12}|^2\right] d\tilde{A}  \label{feLCdim} \\
&+& \int_{\partial \tilde{A}} 
\tilde{W} |(\tilde{Q}_{11},\tilde{Q}_{12})-\tilde{{\bf{q}}}|^2
d\tilde{a}. \nonumber
\end{eqnarray}
Here, there are only two input parameters: the dimensionless elastic constant $\tilde{\kappa}_{el}$ and the dimensionless surface anchoring strength $\tilde{W}$. We drop the tildes in the subsequent text and all results are to be interpreted in terms of the dimensionless variables.

We need to prescribe a suitable form for the reference configuration, $\mathbf{q}$, that enforces the tangent boundary conditions on the edges. Following the formulation in Luo et al.~\cite{Majumdar_2012}, we set ${\bf{q}}(x,y) = s(x,y) (\cos 2\theta(x,y), \sin 2\theta (x,y))$, where $\theta(0,y)=\theta(1,y) = \pi/2$, $\theta(x,0)=\theta(x,1)=0$, and $ s(t,0)=s(t,1)=s(0,t)=s(1,t) = f(t)$. The scalar $s$ has to vanish at the vertices with strong tangent anchoring, because of the mismatch in $\theta$ at the vertices. Hence, we define
\begin{eqnarray}
   f(t) =
\begin{cases}
    t/d ,& 0 \leq t \leq d, \\
    1,   & d \leq t \leq 1-d, \\
    (1-t)/d & 1-d \leq t \leq 1,
\end{cases} \nonumber
\end{eqnarray}
with $d$ chosen to be $3 \sqrt{\kappa_{el}}$. In the liquid crystal literature, $\sqrt{\kappa_{el}}$ is known to be typically proportional to defect core sizes \cite{deGennes_book, Virga_book} and hence, we take $f$ to be approximately unity everywhere (a minimum of the bulk free energy) except for small neighbourhoods around the square vertices, where we expect to see defects.

We can use standard methods from calculus of variations to derive the Neumann boundary conditions on the edges (from the Durand-Nobili surface energy in  Eq.~\eqref{feLCdim}),
\begin{eqnarray}
\frac{\partial Q_{\alpha\beta}}{\partial x} &=& - \frac{W}{\kappa_{el}} (Q_{\alpha\beta} - q_{\alpha\beta}) \,\, {\mathrm{on}} \,\, x=1, \nonumber \\ 
\frac{\partial Q_{\alpha\beta}}{\partial x} &=& + \frac{W}{\kappa_{el}} (Q_{\alpha\beta} - q_{\alpha\beta}) \,\, {\mathrm{on}} \,\, x=0, \nonumber \\
\frac{\partial Q_{\alpha\beta}}{\partial y} &=& - \frac{W}{\kappa_{el}} (Q_{\alpha\beta} - q_{\alpha\beta}) \,\, {\mathrm{on}} \,\, y=1, \nonumber \\
\frac{\partial Q_{\alpha\beta}}{\partial y} &=& + \frac{W}{\kappa_{el}} (Q_{\alpha\beta} - q_{\alpha\beta}) \,\, {\mathrm{on}} \,\, y=0. \nonumber
\end{eqnarray}
We use Greek indices to denote the components of the ${\bf{Q}}$-tensor, $Q_{\alpha\beta} = Q_{11}$ and $Q_{12}$ above.

The numerical methods for the computation of the free energy minima, transition states and minimum energy pathways are described in the appendix.

\section{Minimum Free Energy States}
\label{sec:3}

\begin{figure}
\centering
\includegraphics[scale=1.0,angle=0]{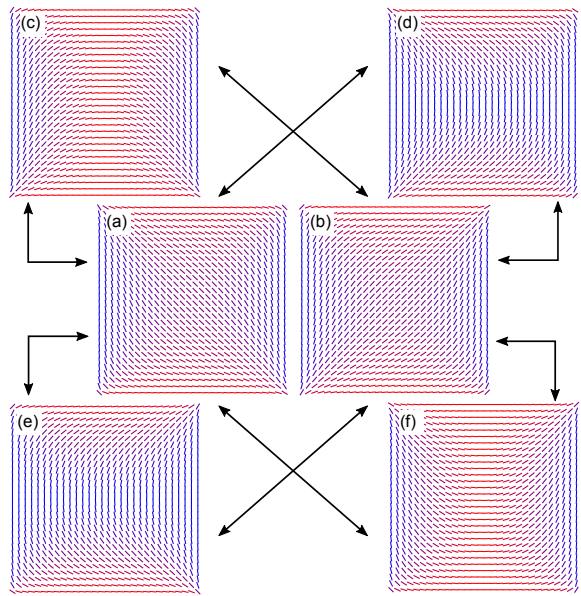}
\caption{Local free energy minima for a square device for $W \ge 1.4 \times 10^{-3}$. There are two classes of configurations: (a-b) diagonal and (c-f) rotated states. Each double arrow indicates the two minima are connected by a single transition state, thus visualising the connectivity of the free energy landscape.}
\label{minima1}
\end{figure}
We first study the minimum free energy morphologies of the system as the surface anchoring $W$ is varied. This is similar to the work reported in \cite{Majumdar_2012} where the authors use continuation methods to compute stable equilibria of this system as $W$ is varied. Throughout this paper, we fix the value of the dimensionless elastic constant $\kappa_{el} = 4 \times 10^{-4}$ (for typical values of the elastic constant and Landau-de Gennes bulk potential parameters, this describes a micron-scale well) and use a square lattice with $150 \times 150$ grid points. 

In agreement with previous studies ~\cite{Majumdar_2012, Aarts_2014,Kusumaatmaja_2014}, we find that for large anchoring strengths i.e. for $W \ge 1.4 \times 10^{-3}$, there are two classes of minima, namely diagonal and rotated states. We check the stability of the solutions by numerical computations of the Hessian of the Landau-de Gennes energy and its eigenvalues at a given solution, as is standard for numerical stability analysis. Due to the system symmetry, there are two equivalent diagonal states and four equivalent rotated states. These states are shown in Fig.~\ref{minima1}.

\begin{figure}
\centering
\includegraphics[scale=1.0,angle=0]{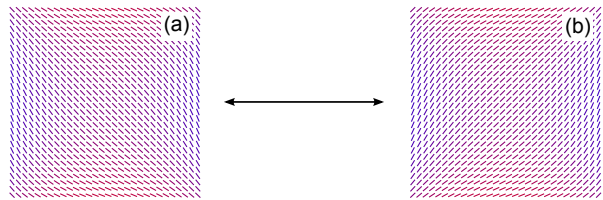}
\caption{Minimum free energy diagonal configurations in a square device for $W < 1.4 \times 10^{-3}$. The double arrow indicates that the two minima are connected by a single transition state.}
\label{minima2}
\end{figure}

For weak anchoring, $W < 1.4 \times 10^{-3}$, the rotated states are no longer stable minima. This is consistent with the bifurcation diagram in \cite{Majumdar_2012} where the authors do not find rotated solutions for small $W$. However, the rotated solutions do survive as transition states in the free energy landscape for weak anchoring, connecting the stable diagonal solutions. This is further explained in the next section and the two stable minima (diagonal states) are shown in Fig.~\ref{minima2}.

Each double arrow in Figs.~\ref{minima1} and~\ref{minima2} indicates that there exists at least one transition pathway between the connected minima and we discuss the transition pathways in the next section.

\section{Transition Pathways}
\label{sec:4}

Previous studies have been limited to free energy minima in the planar bistable device~\cite{Tsakonas_2007,Majumdar_2012}. In this paper we systematically compute the transition states, transition pathways and optimal transition pathways as a function of the surface anchoring parameter $W$. To the best of our knowledge, this has not been reported elsewhere in the literature.

We distinguish three separate regimes depending on the values of $W$: (i) strong anchoring regime, (ii) moderate anchoring regime and (iii) weak anchoring regime.

{\it{Regime I: Strong anchoring regime}} - The first regime is the strong anchoring limit, which we find to occur for $W \ge 6.5 \times 10^{-3}$. The optimal transition pathways are independent of $W$ in this regime whilst they are sensitive to $W$ for weaker anchoring. The computations in this regime are in fact very similar to the Dirichlet case considered in \cite{Kusumaatmaja_2014} where $\mathbf{Q}$ is fixed to be $\mathbf{q}$ on the square edges.

We first make a few comments about the vertex defects in the diagonal and rotated solutions, both of which are local free energy minima in this regime. There are two types of point defects in these solutions: (i) splay defects wherein the director splays outward from the vertex
and (ii) bend defects, for which the director bends between a pair of intersecting edges.

For a diagonal solution, there are two diagonally opposite splay defects and two diagonally opposite bend defects and for a rotated solution, the two splay defects are connected by an edge, as are the two bend defects. For a rotated to diagonal transition, we need two defect transformations along an edge: a bend to splay defect transformation and a splay to bend
defect transformation via director rotation.  We obtain at least three different transition pathways between a rotated and a diagonal solution, as illustrated in Figure~\ref{Strong} below, all of which are mediated by motion of point defects.

\begin{figure}
\centering
\includegraphics[scale=1.0,angle=0]{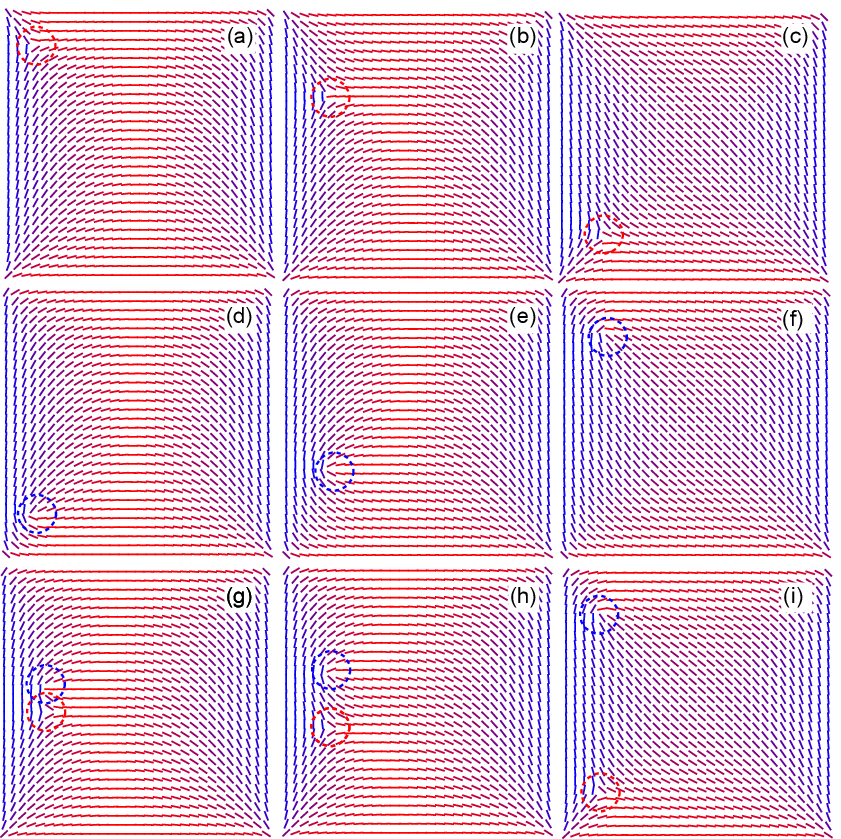}
\caption{Possible transition pathways between rotated and diagonal states in the strong anchoring limit, $W \ge 6.5 \times 10^{-3}$. The transitions are mediated by a $-1/2$ defect, a $+1/2$ defect, and a pair of $\pm 1/2$ defects respectively in panels (a-c), (d-f), and (g-i).}
\label{Strong}
\end{figure}
\begin{figure}
\centering
\includegraphics[scale=1.0,angle=0]{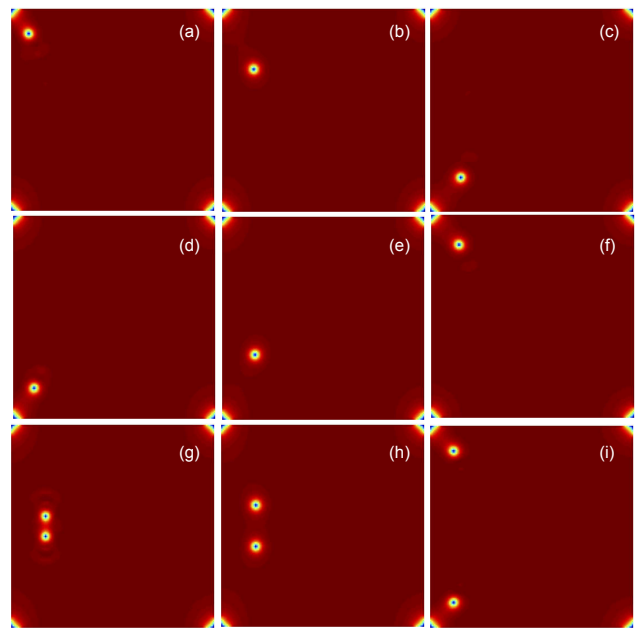}
\caption{(color online) The order parameter plots, $s(x,y)$, for the transition pathways shown in Fig.~\ref{Strong}. $s \rightarrow 0$ at the defect locations and $s \rightarrow 1$ elsewhere. The transitions are mediated by a $-1/2$ defect, a $+1/2$ defect, and a pair of $\pm 1/2$ defects respectively in panels (a-c), (d-f), and (g-i).}
\label{Strong2}
\end{figure}

 The first mechanism, as shown in Fig.~\ref{Strong}(a-c), is mediated by a $-1/2$ defect. The defect is created near the top left corner, propagates towards the bottom left corner and subsequently settles there to yield a bend defect. The numerical results suggest that
the singular bend behaviour at the top left corner of the initial rotated solution is propelled into the square interior leading to the creation of a $-1/2$ defect. As the $-1/2$ defect propagates downwards, it rotates simultaneously and induces a global director rotation within the square domain. Finally, the point defect settles at the bottom left corner to become a bend defect and the system relaxes into a diagonal solution. The second mechanism, Fig. ~\ref{Strong}(d-f), is dominated by a $+1/2$ defect created near the bottom left corner. The rotated solution has a splay defect at the bottom left corner; this defect is propelled into the square interior, it moves upwards along the square edge and whilst moving, rotates and induces a global director rotation. It finally settles at the top left corner as a splay defect and the system relaxes into a diagonal solution. These two mechanisms have identical free energy profiles, as shown in Figure 5. Further analysis using our numerical methods also shows that there can be other transition pathways with higher free energy barriers. We interpret the optimal transition pathway as being the minimum energy pathway with the smallest barrier and the least energetically expensive transition state. An example of a transition pathway with a higher free energy barrier is shown in Fig. ~\ref{Strong} (g-i), where the pathway is mediated by the creation and propagation of a pair of $\pm 1/2$ defects. We also note that the transition pathways  between other rotated and diagonal states can be related by symmetry  to the specific example shown in Fig.~\ref{Strong}. In Fig.~\ref{Strong2}, we plot the order parameter along the transition pathways, as shown in Fig.~\ref{Strong}. The defects at the four corners, along with the bulk $\pm 1/2$ defects (mediating the transitions), are clearly identified by small order i.e. by  small values for $s$ such $s = \sqrt{Q_{11}^2 + Q_{12}^2} \rightarrow 0$ near the defects.
\begin{figure}
\centering
\includegraphics[scale=0.9,angle=0]{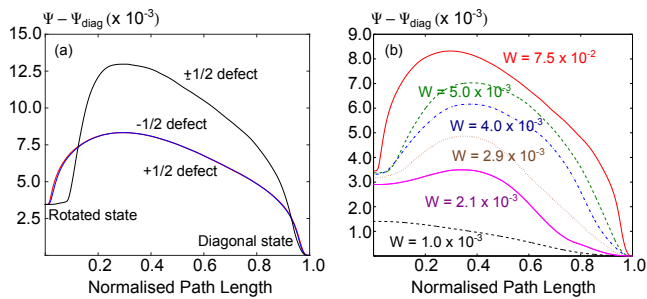}
\caption{(a) The free energy profiles for the three transition mechanisms discussed in Fig.~\ref{Strong}. The $-1/2$ and $+1/2$ defect mechanisms have degenerate free energy profiles, while the $\pm 1/2$ defect pair mechanism has a considerably higher free energy barrier. (b) The free energy
profiles along the optimal transition pathway for different values of the surface anchoring strength $W$.}
\label{EnergyStrong}
\end{figure}

The free energy profiles for the three transition mechanisms discussed in Fig.~\ref{Strong} are shown in Fig.~\ref{EnergyStrong}(a). The $x$-axis corresponds to the ``{\it{normalised path length}}'' between configuration $m$ along the pathway and the initial rotated state,
\begin{equation}
s^m = \frac{1}{N} \left[ \sum_{n=0}^m \sum_{\alpha\beta}\sum_{ij} \left| ({\bf{Q}}_{\alpha\beta}^{ij})^{n+1} - ({\bf{Q}}_{\alpha\beta}^{ij})^n \right|^2 \right]^{1/2}.
\end{equation}
The normalization constant $N$ is chosen such that the total path length between the rotated and diagonal states is 1. For the first two mechanisms described above, the profiles are virtually indistinguishable and the maxima in the free energy profiles correspond to the saddle points (transition states) in the free energy landscape of the system. The transition state configurations are depicted in Figs.~\ref{Strong}(b), (e), and (h). It is worth noting that we do not find any direct pathway between two diagonal or two rotated states.  A transition between two diagonal states or two rotated states is always composed of a sequence of rotated-diagonal transitions. 

%
%
%
%
%
%

{\it{Regime II: Moderate Anchoring}} - We label the range $1.4 \times 10^{-3} \le W < 6.5 \times 10^{-3}$ as the medium anchoring regime. There are three key features of this regime: (i) the diagonal and rotated solutions survive as local free energy minima, (ii) the free energy barrier (along minimum energy pathways) decreases monotonically with $W$ and (iii) the optimal transition pathways do not feature defects.
\begin{figure}
\centering
\includegraphics[scale=1.0,angle=0]{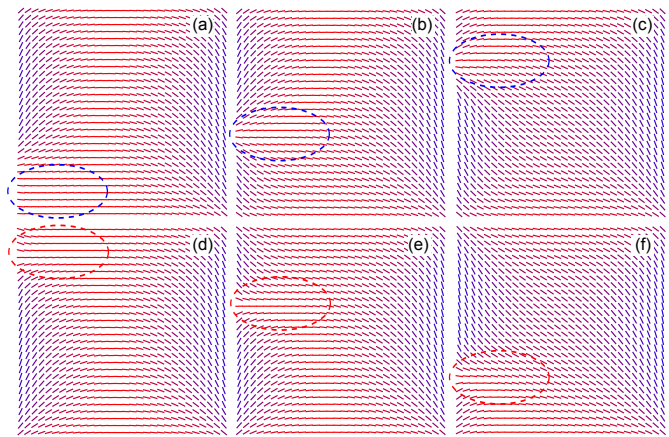}
\caption{Two competing transition pathways in the medium anchoring regime, $2.4 \times 10^{-3} \le W < 6.5 \times 10^{-3}$. The molecules rotate clockwise in the first mechanism (a-c). The director rotation sweeps upward. In the second mechanism (d-f), the rotation is also
clockwise, but it starts from the top left corner and propagates downward. }
\label{Medium}
\end{figure}

This regime is further divided into two sub-regimes (a) $2.4 \times 10^{-3} \le W < 6.5 \times 10^{-3}$ and (b) $1.4 \times 10^{-3} \le W < 2.4 \times 10^{-3}$. For case (a),  the transition from a rotated to a diagonal solution is due to a ``{\it{sequential}}'' director rotation on one side of the square device, without any transient defects. Heuristically, it is energetically preferable to break the tangential anchoring along a square edge leading to director rotation along an edge as opposed to the creation of $\pm 1/2$ defects. As in the strong anchoring regime, we find that there are two mechanisms with degenerate free energy profiles (including the free energy barrier).

In the first mechanism, the director starts to rotate clockwise from the bottom left corner, breaking the tangential anchoring along the left vertical edge. The rotation then propagates upwards, as shown in Fig.~\ref{Medium}(a-c) and this rotation suffices to transform the
splay defect at the bottom left corner into a bend defect and conversely, the bend defect at the top left corner into a splay defect. The final state is a diagonal solution. The second mechanism is shown in Fig.~\ref{Medium}(d-f). The director rotates clockwise from the top
left corner and the sweeping motion propagates downwards, inducing a global director rotation and a transition from a rotated solution to a diagonal solution. The configurations in panels (b) and (e) correspond to the transition states and the two mechanisms have identical free energy profiles.
\begin{figure*}
\centering
\includegraphics[scale=1.0,angle=0]{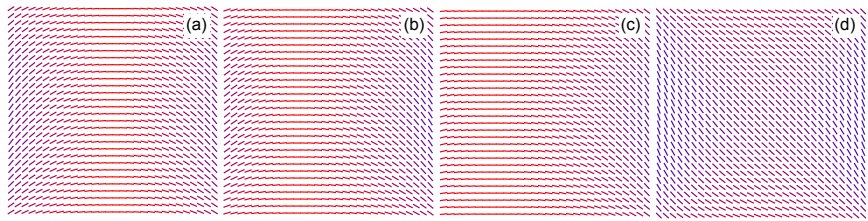}
\caption{The minimum energy transition pathway between a rotated and a diagonal state in the medium anchoring regime, $1.4 \times 10^{-3} \le W < 2.4 \times 10^{-3}$. Panel (b) corresponds to the transition state along this pathway.}
\label{Weak}
\end{figure*}

For weaker anchoring, for $1.4 \times 10^{-3} \le W < 2.4 \times 10^{-3}$, the penalty for breaking tangential anchoring is weak and 
the rotated to diagonal transition follows from a ``{\it{global}}'' director rotation along the left vertical edge. As shown in Fig.~\ref{Weak}(a-d), the director rotates simultaneously along the entire length of the left vertical edge and this rotation induces a global director distortion within the square, leading to a rotated to diagonal transition.

\begin{figure*}
\centering
\includegraphics[scale=1.0,angle=0]{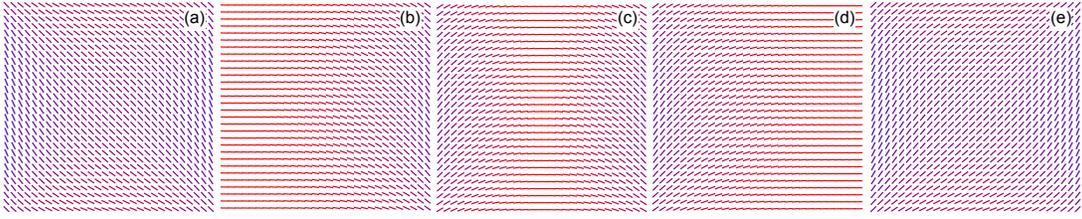}
\caption{One of four equivalent optimal transition pathways between two diagonal states in the very weak anchoring regime, $W < 1.4 \times 10^{-3}$. Panel (c) corresponds to the transition state, which is reminiscent of a rotated state.}
\label{VeryWeak}
\end{figure*}

{\it{Regime III: Weak Anchoring}} - As the surface anchoring strength $W$ further decreases, for $W < 1.4 \times 10^{-3}$, the rotated solutions are no longer stable. The disappearance of the rotated states correspond to cusp catastrophes~\cite{Wales_2001}, where typically a
minimum and two transition states coalesce to a single point. This point, reminiscent of the rotated state (as such, there are four of them), is a saddle point (transition state) between the two stable diagonal solutions.  The transition state in panel Fig.~\ref
{VeryWeak} (c) resembles a rotated solution, with a uniform director profile in the square interior accompanied by two transition layers near the vertical edges. The corresponding free energy profile can be seen from Fig.~\ref{EnergyStrong}(b) for $W = 1.0 \times 10^{-3}$. In this regime, the minimum energy pathways may therefore be represented by the diagram in Fig.~\ref{minima2} where the double arrow represents four equivalent transition pathways.




The movies for all the possible transition pathways described above are available as supporting information.

\section{Conclusions}
\label{sec:5}

We have systematically studied the free energy landscape of a multistable nematic liquid crystal device as we vary the strength of the surface anchoring potential. Previous work focuses on the free energy minima and we  take this work further by computing transition states, 
transition pathways and optimal transition pathways between minima, along with the free energy barriers.

We have classified the system behaviour into three categories, according to surface anchoring strength and qualitative properties of transition pathways.
In the strong anchoring regime, $W \ge 6.5 \times 10^{-3}$, we obtain two distinct classes of free energy minima: diagonal and rotated solutions, as expected from previous work. The optimal transition pathways and the corresponding transition states feature either a $+1/2$ or a $-1/2$ defect, localized near an edge. It is worth noting that these transition states  bear strong resemblance to experimental observations of states with internal defects reported in~\cite{Aarts_2014}, suggesting possible metastability of transition states.  Further, there are multiple transition pathways between free energy minima, which can be mediated by multiple defects but with higher free energy barriers than the optimal pathways.

The existence of a transition state or a saddle point-type critical point of the Landau-de Gennes energy (\ref{feLCdim}), connecting the stable diagonal and rotated solutions in the strong anchoring regime, can be theoretically justified from the celebrated Mountain Pass Theorem \cite{almeida}. In the limit of infinite anchoring, one can define a topological classification scheme for two-dimensional director fields (see \cite{jrmz} for a similar three-dimensional topological classification), for which the diagonal and rotated solutions belong to different topological sectors. This can also be seen more intuitively from the arrangement of the splay and bend point defects in the diagonal and rotated solutions respectively. The Mountain Pass Theorem guarantees the existence of at least one critical point of the Landau-de Gennes energy connecting the topologically distinct diagonal and rotated solutions. In our framework, Mountain Pass solutions correspond to transition states along optimal transition pathways.



In the moderate anchoring regime, for $W < 6.5 \times 10^{-3}$, we do not observe any defect along the optimal transition pathways. The rotated-to-diagonal transition is driven by clockwise director rotation (and anti-clockwise for the opposite diagonal-to-rotated transition) localized along an edge. For $2.4 \times 10^{-3} \le W < 6.5 \times 10^{-3}$, the rotation starts from one corner and propagates to another corner, along an edge. For weaker anchoring strengths ($W < 2.4 \times 0^{-3}$), it is preferable for the director to rotate along an entire edge, since the energy penalty for violating tangent anchoring is relatively weak.

For $W < 1.4 \times 10^{-3}$, we  capture the cusp catastrophe events as the rotated states become unstable and act as transition states for the transition pathways connecting two stable diagonal solutions. This was previously not noted in the literature. 

We note that the same effect i.e. loss of stability of the rotated solutions, can also be achieved by continuously decreasing the square size whilst maintaining strong anchoring on the square edges. In \cite{robinson}, the authors study the critical points of the two-dimensional Landau-de Gennes energy as a function of the square size, with strong anchoring on the edges. For micron-sized wells, they recover the diagonal and rotated solutions as we do. For a critical well size, the rotated solutions lose stability (but exist as non-minimizing solutions) whilst the diagonal solutions retain stability. For nano-scale wells, there is a unique critical point, referred to as the order-reconstruction solution, by analogy with similar findings in \cite{Kralj_2014}.

For moderate and weak anchoring, the height of the free energy barrier decreases monotonically with $W$ whilst the height of the free energy barrier is independent of $W$ in the strong anchoring regime. Secondly, the configurations of the transition states and the free energy barriers depend strongly on the anchoring strength, for moderate and weak anchoring. These results can therefore be exploited to fine-tune the desirable surface properties i.e. use an anchoring strength that yields the optimal free energy barrier for device applications. For example, a very low barrier compromises the stability of the minimum free energy configurations and the device performance. A very high barrier is an impedance to realistic switching, necessitating large power input or strong external electric fields.

We will now discuss how the dimensionless free energy can be converted to SI unit (Joules). Note that the free energy defined in Eq.\eqref{feLCdim} is in two dimensions (i.e. free energy per unit thickness), while all practical experiments are, of course, in three dimensions. As such, we need to account for the thickness of the device, $\Delta z$, such that $\Psi = (\alpha^2L^2 \Delta z /C) \tilde{\Psi}$, where $\tilde{\Psi}$ is the dimensionless free energy defined in Eq. \eqref{feLCdim}. The parameters $\alpha$ and $C$ are experimentally measurable quantities. As an example, for the common liquid crystal material MBBA \cite{Mkaddem_2000}, the characteristic bulk constants are of order $\alpha \sim 4.2 \times 10^2 \, \rm{J/m^3}$ and $C \sim 3.5 \times 10^3 \, \rm{J/m^3}$. To estimate $\alpha$, we have assumed that the temperature is $1^\circ$ below the critical nematic transition temperature ($\sim 46^\circ$C ). From Fig. \ref{EnergyStrong}, the typical free energy barrier in dimensionless units is of order $10^{-3}$. Assuming the thickness of the device is 1 micron and the area of the device is (100 micron)$^2$, the typical free energy barrier in SI unit is therefore of order $5.1 \times 10^{-16}$ J. This is, of course, much larger than $k_bT \sim 4.4 \times 10^{-21} J$. On the other hand, if the device dimension can be miniaturized to 1 micron x 1 micron x 100 nm, then the typical free energy barrier becomes of order $5.1 \times 10^{-21}$ J, comparable to $k_bT$. The free energy barriers for other materials and device dimensions can be computed in an analogous way.

Our methods can be extended to three dimensions and to include external fields. For example, we could use similar methods to study transition pathways in three-dimensional set-ups, as studied in  \cite{Kralj_2014}, where the authors employ a three-dimensional Landau-de Gennes modelling approach and find a new biaxial order-reconstruction pattern for shallow nano-scale rectangular wells with strong anchoring. We speculate that the biaxial order-reconstruction pattern loses stability for larger wells but exists as a transition state connecting diagonal and rotated solutions for large micron-scale wells with strong anchoring.

Similarly, we can use these numerical methods to describe more complex geometries and boundary conditions, for example the Zenithally Bistable Device studied in \cite{Majumdar_2014}. They are also compatible with other simulation techniques tailored to the dynamics of liquid crystalline systems or generally complex fluids problems (such as finite element, lattice Boltzmann method, etc). These numerical methods promise to be a valuable design or optimization tool, allowing us to obtain a detailed picture of the free energy pathways of a complex system, and subsequently how the pathways evolves as the system parameters are varied.


\section{Appendix: Computational Techniques}

Here we summarise the most salient features of the computational techniques. A more detailed description and implementation of the methods can be found in~\cite{Kusumaatmaja_2014}.

\subsection{Discretization of the Free Energy Functional} 
We discretize the Landau-de Gennes free energy on a square grid. Specific stencils are needed to approximate the derivatives and it is important that they are at least second order accurate. We have used
\begin{eqnarray}
|\nabla Q^{ij}_{\alpha\beta}|^2 &=& \frac{\left[ ( Q^{(i+1)j}_{\alpha\beta}-Q^{ij}_{\alpha\beta})^2 + (Q^{(i-1)j}_{\alpha\beta}-Q^{ij}_{\alpha\beta})^2 \right]}{2(\Delta{}x)^2} \nonumber \\
&+& \frac{\left[ ( Q^{i(j+1)}_{\alpha\beta}-Q^{ij}_{\alpha\beta})^2 + (Q^{i(j-1)}_{\alpha\beta}-Q^{ij}_{\alpha\beta})^2 \right]}{2(\Delta{}y)^2}, \nonumber
\end{eqnarray}
where the superscripts $i$ and $j$ label the lattice points in two dimensions. We have used a 150x150 grid based on the results of Luo et al. \cite{Majumdar_2012}, where they carried out convergence tests on the energy-minimizing solutions in the Landau-de Gennes framework.

{\it{Minimum Free Energy States}} - The first step in a survey of the free energy landscape is to compute the majority, if not all, of the possible minima in the system. We follow a stochastic approach, where each step consists of a trial move followed by an energy minimization. It is identical to a basin-hopping algorithm~\cite{Li_1987,Wales_1997} at
infinite temperature, such that every minimum state found is recorded. The simplest trial move consists of random perturbations of the lattice field values, $(Q')^{ij}_{\alpha\beta}= Q^{ij}_{\alpha\beta} + \xi \Theta$. Here $\xi$ is a random number between -1.0 and 1.0, and $\Theta$ is the amplitude of the perturbation, which we usually take to be $\Theta = 0.5$. The energy minimization is carried out using the limited-memory Broyden-Fletcher-Goldfarb-Shanno (LBFGS) algorithm~\cite{Nocedal_1980,Liu_1989}. In this paper, we typically take 500 (basin-hopping) steps to sample the minima of the system.

\subsection{Transition States and Minimum Energy Pathways} 
Given the minimum free energy states, we numerically compute the transition states (saddle points in the free energy landscapes) using a combination of the doubly-nudged elastic band (DNEB) method~\cite{Trygubenko_2004, Ruhle_2013} and the hybrid eigenvector-following technique~\cite{Wales_book}. Transition states are special saddle points or critical points in the free energy landscapes where the energy gradients are zero in all eigendirections and one of the eigenvalues of the Hessian is negative. This is to be contrasted to minima, where the gradients are zero and the eigenvalues are positive in all eigendirections. 

The DNEB method is a double-ended search algorithm for finding transition states and minimum energy pathways between any two pair of minima. Minimum energy pathways have an important feature such that every point on such a pathway is a minimum in all but a single distinguished direction in the phase space. As a consequence, a maximum along the minimum energy pathway corresponds to a transition state.

In DNEB, a  set of images, $\{ {\Gamma}^1, {\Gamma}^2,\ldots, {\Gamma}^N\}$, are placed between the two endpoints (minima), ${\Gamma}^0$ and ${\Gamma}^{(N+1)}$. The symbol ${\Gamma}$ represents all the degrees of freedom in the system, i.e. $\{Q_{\alpha\beta}^{ij}\}$. We typically use 30 images, and they are initialised by taking a linear interpolation between the two endpoints.

The images are relaxed using Landau-de Gennes energy gradients that include contributions from two components. The first contribution, $\bf{g}$, comes from the derivatives of the free energy functional with respect to the lattice degrees of freedom, $d\Psi/dQ_{\alpha\beta}^{ij}$. The second contribution is a spring force, $\tilde{\bf{g}}$, which is required to keep the images roughly equidistant. The following spring potential is applied on every image
\begin{equation}
V^\alpha_{\rm{spring}} = \frac{k}{2} \left( (s^{\alpha,-})^2 - (s^{\alpha,+})^2 \right).
\end{equation}
$s^{\alpha,-}$ and $s^{\alpha,+}$ are respectively the distances to the left and right images, $(s^{\alpha,\pm})^2 = |{\Gamma}^{\alpha\pm1}-{\Gamma}^\alpha|^2$. The coefficient $k$ is the spring constant.

Using the true and spring gradients, without any projection, often results in two issues~\cite{Henkelman_climbing_2000,Henkelman_improved_2000}: (i) corner-cutting, where images are pulled away from the minimum energy path; and (ii) sliding-down problems, where images slide down from barrier regions. In DNEB, we use two projections. First, we only retain the components of the true gradient that are perpendicular to the unit tangent vector $\bf{\hat{\tau}}^\alpha$,
\begin{equation}
{\bf{g}}_\perp^\alpha = {\bf{g}}^\alpha - ({\bf{g}}^\alpha\cdot{\bf{\hat{\tau}}}^\alpha) {\bf{\hat{\tau}}}^\alpha. \label{realgrad}
\end{equation}
Secondly, the following component of the spring constant gradient is retained,
\begin{equation}
{\bf{\tilde{g}}}_{\rm{DNEB}}^\alpha = k (s^{\alpha,+} -s^{\alpha,-}){\bf{\hat{\tau}}}^\alpha + {\bf{\tilde{g}}}_{\perp}^\alpha-({\bf{\tilde{g}}}_{\perp}^\alpha\cdot{\bf{\hat{g}}_{\perp}}^\alpha){\bf{\tilde{g}}}_{\perp}^\alpha, \label{springgrad}
\end{equation}
where ${\bf{\tilde{g}}}_{\perp}^\alpha = {\bf{\tilde{g}}}^\alpha - ({\bf{\tilde{g}}}^\alpha\cdot {\bf{\hat{\tau}}}^\alpha){\bf{\hat{\tau}}}^\alpha$. The unit tangent ${\bf{\hat{\tau}}}$ is defined as in~\cite{Henkelman_improved_2000}. As a convergence criterion, we stop the DNEB run when the root mean square of the gradients as defined in Eqs. \eqref{realgrad} and \eqref{springgrad} in the manuscript is $< 10^{-9}$, which in our experience is more than enough to ensure the calculations have converged.

 A maximum in the DNEB path corresponds to a transition state candidate, which we then refine by applying the hybrid eigenvector-following technique. We use a Rayleigh-Ritz approach~\cite{Wales_book}, based on gradients of the Landau-de Gennes energy, to compute the transition states since this method is more efficient for large systems. 

Once the transition states are found, small displacements are applied in the two downhill directions. Energy minimizations are again carried out using the LBFGS algorithm~\cite{Nocedal_1980,Liu_1989}. Given a minima -  transition state - minima triplet, this run yields the transition pathway for this triplet and the corresponding free energy barrier. We check that this pathway is consistent with the one obtained from DNEB calculations. The free energy barriers could be interpreted as the free energy differences between the transition state and the free energy minima. There are typically multiple transition pathways between a pair of free energy minima. The optimal pathway is the pathway with the smallest free energy barrier.





\footnotesize{
\bibliography{multistable} 
\bibliographystyle{rsc} 
}

\end{document}